# Robust Pricing with Refunds[*]

Toomas Hinnosaar[†]     Keiichi Kawai[‡]

**Abstract**

Before purchase, a buyer of an experience good learns about the product's fit using various information sources, including some of which the seller may be unaware of. The buyer, however, can conclusively learn the fit only after purchasing and trying out the product. We show that the seller can use a simple mechanism to best take advantage of the buyer's post-purchase learning to maximize his guaranteed-profit. We show that this mechanism combines a generous refund, which performs well when the buyer is relatively informed, with non-refundable random discounts, which work well when the buyer is relatively uninformed.

## 1 Introduction

A buyer of an experience good often learns something about a product's match to her preferences through various sources before purchase. The seller, however, may not know exactly what information a buyer has acquired, how a buyer gathers information or how she processes it. For instance, an online shoe retailer may be unaware of the level of "showrooming" conducted by a shopper prior to visiting his website, and is thus unable to discern if the buyer already knows about whether a pair of shoes will fit. Similarly, a booking agent does not know if a traveler planning to book a flight already has a specific travel itinerary in mind or whether the individual already knows about what particular in-flight experience she wants. In both instances, uncertainty about the buyer's prior information limits the seller's ability to extract profits from the buyer should a trade occur.

[*]We are grateful to Mark Armstrong, Pak Hung Au, Gary Biglaiser, Ralph Boleslavsky, Laura Doval, Juan Dubra, Jeff Ely, Dino Gerardi, Paolo Ghirardato, Marit Hinnosaar, Johannes Hörner, Atsushi Kajii, Teddy Kim, Anton Kolotilin, Fei Li, Hongyi Li, Giorgio Martini, Claudio Mezzetti, Ignacio Monzón, Alessandro Pavan, Karl Schlag, Roland Strausz, Balázs Szentes, Alexander Wolitzky, and two anonymous referees for the useful discussions and valuable comments. We would also like to thank seminar participants at SAET (Taipei), the International Conference on Economic Theory and Applications (Chengdu), the International Conference on Game Theory (Stony Brook), Midwest Economic Theory Conference (Nashville), Peking University, Durham Economic Theory Conference, Kyoto University, University of North Carolina at Chapel Hill and Duke University, University of Miami, University of Queensland, and University of Melbourne. We thank Chris Teh for his excellent research assistance. The second author greatly acknowledges the financial support from Australian Research Council (DECRA Grant RG160734), CLMR summar research grant, and UNSW Business School.

[†]Collegio Carlo Alberto and CEPR, `toomas@hinnosaar.net`.
[‡]UNSW, Sydney, `k.kawai@unsw.edu.au`.

However, it is often the case that a buyer conclusively learns about the product's fit only after the purchase has been made. For instance, a buyer of an airline ticket may not immediately know her exact travel plans at the moment of purchase one or two months prior. Additionally, she may only develop a much better understanding of the expected itinerary, and therefore the "fit" of her ticket purchase, only a few days prior to flying. This "learning through purchase" feature of experience goods provides an opportunity for the seller to combat uncertainty regarding the buyer's information sources through a refund policy. From the buyer's perspective, a generous refund policy allows her to try the good without taking on huge risk. Through this, the seller is able to reduce the importance of the buyer's prior information on her purchasing decision, in turn allowing him to charge a high price without sacrificing the probability of trade.

At the same time, the seller needs to be wary of costs associated with product return. For each returned product, the seller incurs a restocking cost. More importantly, the seller is often unsure about how informed/uninformed the buyer is prior to purchase, and is therefore unable to evaluate the buyer's possibility of product return at the time of purchase. Because of this, a seller including a generous provision for product returns in the hopes of earning higher profits by enticing potential buyers to purchase the product at a higher price may suffer a significant loss if the buyer turns out to be the one likely to return the product. As such, the seller needs to set a refund policy that insures himself from these potentially unprofitable outcomes.

We therefore ask the following questions. When faced with uncertainty about the prior information that a potential buyer might have about the product, does the seller always benefit from offering a refund policy? If so, what is the best way to utilize the buyer's "learning through purchase" to combat uncertainty? Our answer is yes to the first question, and, to answer the second, we show that among all possible mechanisms, a simple mechanism that combines a generous refund policy with random non-refundable discounts allows the seller to achieve the best guaranteed-profit.

To answer these questions, we analyze a bilateral trade model, where the buyer's valuation of the product may either be high or low. Whereas this value is initially unknown to both the seller and the buyer, both parties share a common prior belief about the buyer's valuation. The buyer also observes a private signal about her valuation prior to her purchasing decision, but is only able to learn of her true valuation for the product after a purchase has been made.

The first of the two key features of the model is that the seller and the buyer may continue to interact even after the initial transaction has concluded. For instance, the seller can choose whether to allow the buyer to return the product and how much to refund should the buyer choose to do so. In this case, the seller incurs a restocking cost for each returned product. More generally, the seller may choose to employ any direct or indirect mechanism. For example, the seller may choose to use what we call a *pricing policy*—a simple mechanism that randomizes over price-refund pairs. Alternatively, the seller may opt to use a dynamic sequential mechanism that screens the buyer based on the signal distribution, the realized signal, and, post-purchase, her private valuation.

The second key feature of the model is the seller's uncertainty about the buyer's information structure (i.e., the distribution of signals). Whereas traditional Bayesian models assume that the distribution of signals is commonly known,[1] we analyze an environment where the seller neither knows the buyer's signal distribution nor has a prior belief about possible signal distributions, yet wishes to maximize a guaranteed-profit under this partial information.[2]

We adopt the standard approach in robust pricing literature and look for a solution that works well under all possible scenarios according to *the guarantee criterion*, also known as *the maxmin criterion*.[3] Specifically, our goal is to identify two objects: the *best guaranteed-profit*, i.e., the sharp lower bound of the seller's profit with respect to the buyer's possible signal distributions; and a selling mechanism that provides the best guaranteed-profit. We show that such a mechanism combines a generous refund, which performs well when the buyer is relatively informed, with non-refundable random discounts, which work well when the buyer is relatively uninformed.

To understand the intuition and the implication of the results from our robustness analysis, we start with the observation that the aforementioned pricing policy provides some intuitive guarantee to the seller who only has partial information about what the buyer knows. That is, the seller can partially hedge against uncertainty through pricing policies. The rationale behind this is as follows. We start by considering when the seller sets a single price with certainty. A buyer who is relatively uninformed would face little variation in the signal that she receives, and as such have a high likelihood of receiving a moderate signal. Setting a single high price would likely result in no sales. Alternatively, it is possible for the buyer to be relatively informed such that there is large signal variation. Given this, the signal received by the buyer can be either highly favorable or unfavorable. In this case, setting a low price leaves "money on the table", as the buyer with a favorable signal would be willing to pay more, whereas those with an unfavorable signal would not purchase the product anyway.

Considering both cases, the seller can guarantee a better profit by randomizing the price he sets. Through this, the seller increases the chance of trading with an uninformed buyer without sacrificing too much of a profit from selling to the buyer who has received a favorable signal. The randomization of prices that maximizes the seller's worst-case profit with respect to the buyer's signal, which we refer to as *robust random pricing*, is known to be log-uniformly distributed prices on a certain range.[4]

Alternatively, a seller may improve upon his guaranteed-profit when setting a single price by offering a refund upon product return. A generous refund reduces the importance of the buyer's signal in the purchasing decision and thus allows the seller to increase the price he sets, in turn increasing his revenue from the initial transaction. However, depending on the buyer's signal distribution, offering a refund may significantly increase the likelihood of product return, in turn increasing the likelihood of restocking costs being borne by the seller. To make matters worse, the seller has no prior knowledge of the buyer's underlying signal distribution nor the signal received by the buyer, and thus cannot evaluate the likelihood of product return when deciding on the refund amount provided. Therefore, the seller must design a refundable offer such that the offer is accepted by the buyer if and only if she is unlikely to make a product return, i.e., the buyer receiving a sufficiently favorable signal. The seller can achieve this by constructing a price-refund pair such that the marginal buyer – i.e., the buyer who is indifferent between accepting and rejecting the offer – provides the seller with zero expected profits if she accepts the offer. Of all the possible price-refund pairs that satisfy this property, we refer to the pair in which the price and refund is set at the buyer's highest possible valuation as that of *the generous refund*.

---

[1] See for example, Inderst and Tirosh (2015) and Krähmer and Strausz (2015).

[2] A possible modeling choice to capture such uncertainty is to introduce a hierarchy of higher-order beliefs about the buyer's possible information structure. However, as Savage (1972) puts it, "Such a hierarchy seems very difficult to interpret, and it seems at best to make the theory less realistic, not more."

[3] This approach was first suggested in decision theory by Savage (1972) and axiomatized by Gilboa and Schmeidler (1989). For a recent literature review on robust mechanism design and contracting, see Carroll (2018).

[4] This environment is a special case of Du (2018).

Intuitively, the seller should be able to further hedge against uncertainty and receive a better guaranteed profit by combining both tools. More specifically, consider *the robust refund policy*, that is, a pricing policy that randomizes over the generous refund and *random discounting*. Under *the robust refund policy*, lower prices appear more frequently in comparison to that of robust random pricing.[5] Relative to robust random pricing, random discounting provides the seller with greater profits when the buyer is relatively uninformed, but lesser profits when the buyer is relatively informed. On the other hand, the generous refund by design results in a large profit when the buyer receives a favorable signal, and no profit otherwise. In this sense, it performs well when the buyer is relatively informed and poorly when the buyer is relatively uninformed. Thus, we would expect that a mixture of the generous refund (that is only attractive to the buyer with a sufficiently favorable signal) and random discounting (that is attractive to the buyer only when the signal she receives is moderate or favorable) provides the seller with a better guaranteed-profit than robust random pricing or the generous refund alone.

Our main result formally shows that this intuition is correct: *the robust refund policy*, a simple mechanism that combines the generous refund and random discounting, indeed provides the best guaranteed-profit among all other possible mechanisms, including that of more complicated sequential screening mechanisms. Furthermore, the robust refund policy satisfies the guarantee criterion and thus distinguishes itself from other possible simple mechanisms, addressing Wilson's critique (Wilson (1987)) by being a simple mechanism with good performance over a wide range of possible buyer's signal distributions.[6]

Our robustness analysis also identifies a *worst-case distribution*, i.e., a distribution of the buyer's signal that minimizes the highest profit of the seller even in the case in which the seller can observe the buyer's true signal distribution. Our result demonstrates that the seller's profit in the case of the worst-case distribution is bounded from above by the best guaranteed-profit even when the seller knows the distribution. In this sense, the seller uses the robust refund policy because he cannot exclude the possibility that the buyer's signal distribution is the worst-case distribution. We then naturally pose the following question: When, if ever, should the seller be worried about such a worst-case signal distribution?

---

[5]The policy randomizes over price-refund pairs prior to the buyer's purchasing decision. The buyer only observes the single realized price-refund pair when deciding whether to buy the product.

[6]We analyze an environment where the interaction between the seller and the buyer may continue after the initial transaction, which itself is a part of the seller's mechanism design problem. Thus, nothing guarantees a robust mechanism to be simple.

The second contribution of our article is to provide an answer to this question. We show that the worst-case distribution arises as an equilibrium of a game in which the buyer first chooses a signal distribution, and the seller responds by designing a selling mechanism.[7] In this sense, the seller will want to be concerned of the worst-case distribution if (but not only if) he cannot exclude the possibility that the buyer can choose how much to learn and believes that the seller can observe her signal distribution.

Our findings offer a novel rationale for generous return policies: they hedge the seller against uncertainty in situations where the seller is unsure about how much prior information a buyer may have about the product match before the purchase decision. The literature has identified various other reasons that explain why companies use return policies: e.g., as costly signals for product quality and product fit for the consumer (Grossman (1981); Moorthy and Srinivasan (1995); Inderst and Ottaviani (2013)), as insurance for risk-averse consumers (Che (1996)), and as a tool for price discrimination (Zhang (2013); Escobari and Jindapon (2014); Inderst and Tirosh (2015)).[8]

Among these, the closest to our article is that of Inderst and Tirosh (2015). In an environment where the seller knows the buyer's signal distribution, Inderst and Tirosh (2015) show that return policies work as "metering devices," where refunds make different consumers more similar and thus allow the firm to capture more of the surplus by raising prices.[9] Consequently, the seller sets the refund amount above the restocking cost. In contrast to our article, the authors predict that the optimal refund is interior, i.e., strictly between the salvage value and the price of the product, and thus fail to explain why the seller would promise to refund the original purchase price minus a small fee charged to the buyer, as often observed in practice.

Our article belongs to the growing literature on information design in consumer markets, which has recently shed new light on advertising (Anderson and Renault (2006); Boleslavsky et al. (2019)), consumer search (Anderson and Renault (2006); Armstrong and Zhou (2016); Choi et al. (2019)), price discrimination (Bergemann et al. (2015)), pricing of attributes (Smolin (2019)), and price competition (Armstrong and Zhou (2019)). Among others, our findings are closely related to the results presented in Roesler and Szentes (2017) and Du (2018). Roesler and Szentes (2017) identify the information structure that maximizes the buyer's welfare when the seller best responds to the information structure via uniform pricing. Du (2018) shows that the information structure found in Roesler and Szentes (2017) minimizes the profit that the seller can obtain. Instead, the seller obtains the best guaranteed-profit by what he calls exponential pricing.[10]

The key difference between our article and previous works is that we study situations in which the interaction between the seller and the buyer can continue even after the purchase has been made. Without post-purchase interactions, the buyer's signal affects her purchasing decision only through altering her expected private value conditional on the signal received. If the buyer-seller interaction continues after the buyer's purchase, the buyer also cares about the likelihood of returning the product, which itself depends on the mechanism offered by the seller. Thus, in our model, there exists an interplay between the buyer's signal distribution, the buyer's purchasing decision and the mechanisms the seller can offer.[11] In our model, the seller can indirectly control the impact that buyer's prior information has on her purchasing decision by incentivizing the buyer to learn through purchase.[12] The seller also can sequentially screen the buyer, first by her signal distribution, then by her realized signal; and finally by her realized valuation for the product.[13]

---

[7] For example, the buyer can delegate information gathering to a third party, such as an algorithm or an employee, to commit to a particular information structure.

[8] Escobari and Jindapon (2014) also provide empirical evidence that a fully refundable airplane ticket is typically about 50% more expensive than a non-refundable ticket. However, the difference disappears in the last week before departure. These facts fit well with our model's predictions.

[9] Similar ideas have been studied in other contexts, such as overbooking by airlines, e.g., Ely et al. (2017).

[10] Libgober and Mu (2018) analyze a robust dynamic pricing problem where the product is durable, and buyers learn about their value for the product over time.

Lastly, our model can be interpreted as a game between the seller, who maximizes his profit by indirectly controlling the buyer's learning on product fit through the design of a price-refund pair; and a player called Nature, who minimizes the seller's profit by directly choosing the buyer's signal distribution. In this sense, the game is akin to Bayesian persuasion games with competing information designers. We therefore utilize the concavification technique (Aumann et al. (1995); Kamenica and Gentzkow (2011)) and the properties of equilibrium payoff functions in competitive Bayesian-persuasion settings (Boleslavsky and Cotton (2018); Au and Kawai (2019a,b)) to guide our analysis.

## 2 Model

There is a risk-neutral (male) seller of a product, and a risk-neutral (female) buyer whose valuation for the product is $v \in \{0, 1\}$, i.e., the product either does not fit (with value equal to 0) or does fit (with value normalized to 1). The buyer's valuation $v$ follows a commonly known distribution such that $\mu = \Pr(v = 1)$. Initially, neither the seller nor the buyer know the realization of $v$. The seller's production cost is zero. That is, we assume that the trade between the seller and the buyer is socially efficient.

The interaction between the seller and the buyer takes place over three stages. In the first stage, the buyer receives a signal about the value $v$. In the second stage, the seller offers and commits to a selling mechanism without knowledge of the signal or the underlying stochastic process that generated the signal. In the third stage, the buyer and the seller interact according to the previously specified selling mechanism.

If the mechanism does not allocate the product to the buyer, then the buyer-seller interaction ends. On the other hand, if the mechanism allocates the product to the buyer, then she learns of the realized value $v$. Depending on the specified mechanism, she can keep the product or return it to the seller. We assume that when the buyer returns the product to the seller, the value of the seller's outside option decreases by $c$, which we call the *restocking cost*. We interpret this restocking cost as representing the losses absorbed by the seller when the product is returned. This captures costs such as those associated with processing returns, repackaging and restocking merchandise, product tests, and being unable to re-sell the product again as new. Furthermore, as production costs are assumed to be zero, one may interpret negative salvage value as the returned product being less valuable than that of a newly produced one. Without loss of generality, we further assume that the seller fully covers the buyer's costs of returns, such that the buyer can return the product without incurring any cost.[14]

---

[11]When goods are search goods, information plays a similar role as analyzed in Choi et al. (2019). Given a signal, whether a consumer visits a seller depends not only on the conditional expected value of the good, but also on the likelihood of purchase.

[12]Unlike many articles in the literature on sequential screening and dynamic mechanism design where an uninformed-consumer can become informed over time, the buyer in our model does not learn anything new after the initial purchasing decision should she decide not to buy the product.

[13]The literature on sequential screening and dynamic mechanism design has identified how and why advance sales to still-uninformed consumers can help the seller. See, e.g., Gale and Holmes (1992, 1993); Courty and Li (2000); Eső and Szentes (2007); Nocke et al. (2011); Gallego and Sahin (2010); Ely et al. (2017); and von Wangenheim (2017).

Without loss in generality, we can represent a generic buyer's signal as a posterior $q = \Pr(v = 1)$, a random variable drawn from a cumulative distribution function $F \in \mathcal{F} \equiv \{F : \mathbb{E}_F[q] = \mu\}$. For this reason, we use a distribution $F$ over posteriors to represent the buyer's information structure and call it *a signal distribution*. Analogously, we call the realized posterior $q$ a *signal*.

It is useful to observe that $F \in \mathcal{F}$ if and only if $\int_0^1 F(q)dq = 1 - \mu$. Intuitively, the buyer is relatively informed if the graph of $F$ is relatively flat, such that the likelihood of the posterior being favorable (close to 1) or unfavorable (close to 0) is relatively high. In contrast, the buyer is relatively uninformed if $F$ has a steep slope around the prior $\mu$ and thus the posterior is likely to be moderate (close to $\mu$).

The seller commits to a (direct or indirect) mechanism $M \in \mathcal{M}$ that specifies messages, transfers, and the allocation in two stages. The mechanism must be consistent with the primitives of the model, and incentive-compatible and individually-rational for the buyer in both stages. Other than that, we do not impose any restrictions about the mechanism that the seller can offer. For example, $\mathcal{M}$ contains a (direct) mechanism such that the seller asks the buyer to report his signal distribution $F$ first, the realization of signal $q$, and then the realized value $v$. The set $\mathcal{M}$ also includes randomization over such mechanisms.

However, of particular interest to us is a class of simple mechanisms that consists of randomization over price-refund pairs. More specifically, by an *offer*, and a *pricing policy*, we refer to a price-refund pair $(p, r)$ and a randomization over $(p, r)$ respectively. We use $\mathcal{P} \subset \mathcal{M}$ to denote the set of all possible pricing policies. If the seller uses a pricing policy, then the buyer faces at most two choices. After observing the realized offer $(p, r)$, the buyer decides whether to buy at a price $p$. If the buyer decides not to buy, or buys when a refund is not offered ($r = 0$), then the interaction between the buyer and the seller ends. If the buyer buys the product when a refund is offered ($r > 0$), the buyer learns of $v$ and then decides whether to return the product and receive the refund $r$.

We use $V(M|F)$ to denote the seller's expected profit from mechanism $M \in \mathcal{M}$ when the buyer's signal distribution is $F \in \mathcal{F}$. We say that the mechanism $M$ provides a *guaranteed profit* of $\widehat{V}$ when $V(M|F) \geq \widehat{V}$ for all $F \in \mathcal{F}$, with equality for some $F \in \mathcal{F}$. Our goal is to identify the *best guaranteed-profit* as well as a *mechanism $M$* that provides the *best guaranteed-profit*.[15] To do this, we recast the seller's problem as a zero-sum game between the seller and an (adversarial) player called Nature whose objective is to minimize the seller's profit. More specifically, the seller first chooses a (possibly stochastic) mechanism $M \in \mathcal{M}$. Nature then chooses a signal distribution $F \in \mathcal{F}$ after observing the seller's choice of $M$. The buyer then learns of the signal $q$, and the buyer and the seller subsequently

---

[14]One can interpret our restocking cost $c$ as the sum of the buyer's return costs $c_b$ and the restocking cost of the seller $c_s$ (i.e. $c = c_s + c_b$). If we adjust the (expected) monetary transfer from the seller to the buyer by $c_b$, then our model becomes equivalent to the model in which the buyer bears the cost of return $c_b$.

[15]Formally, the best guaranteed-profit and the set of mechanisms that provide the best guaranteed-profit are $\sup_{M \in \mathcal{M}} \min_{F \in \mathcal{F}} V(M|F)$ and $\arg\sup_{M \in \mathcal{M}} \min_{F \in \mathcal{F}} V(M|F)$.

proceed to interact according to the rules of the mechanism $M$.[16]

In particular, a pricing policy is an indirect mechanism that allocates the product randomly. If the seller were only allowed to use pricing policies, the game proceeds as follows:

1. The seller chooses a pricing policy $P \in \mathcal{P}$.

2. After observing the seller's choice $P$, Nature chooses a signal distribution $F \in \mathcal{F}$.

3. The buyer learns of the offer $(p, r) \sim P$ and the realized signal $q \sim F$, and then decides whether to purchase the product at the price $p$.

4. If the buyer does not buy the product, the game ends and the seller's profit and the buyer's payoff are both zero. If the buyer purchases the product, she learns of the value $v$. The buyer is allowed to return the product for a refund of $r$ if and only if $r > 0$. If the buyer does not return the product, then the seller's profit is $p$ and the buyer's payoff is $v - p$. On the other hand, if the buyer returns the product, the seller's profit is $p - r - c$ and the buyer's payoff is $r - p$.

## Remarks

We now address several key assumptions and interpretations of our model. Firstly, the assumption that $\mu$ is known captures situation in which the seller only has partial information about the aggregate distribution of buyer valuations within society. In particular, we assume that the seller knows the first moment of the population.[17] However, as it is often the case in reality, we assume that the seller cannot observe the sources of information used by a particular buyer and is thus forced to design a mechanism that accounts for this uncertainty.[18] For example, online retailers such as Amazon are often able to estimate $\mu$ through compiling statistics on prior purchasing consumers and through market research. These retailers sell a wide range of experience goods such as clothing and shoes, the value of which to consumers can be highly heterogeneous depending on their prior information sources.

Secondly, we note that our model captures various buyer-seller interactions in which the buyer learns about the product's fit only after making the purchase, even in the case that the product is not an experience good. For instance, a buyer of an airline ticket may not know the details of her itinerary at the time of purchase. After she has a ticket booked, she may refine her traveling plan and come to learn a few days before the flight that the ticket purchased is no longer an ideal match for her. A fully refundable ticket allows the buyer to cancel the flight without incurring significant expenses in such situations. In this context, our model investigates whether and why the airline company should offer the buyer the ability to receive a refund on her ticket should she "change her mind" and increase its guaranteed profit.

---

[16]A (direct) mechanism $M$ can specify the probability that the product being allocated to the buyer based on her report. In such an instance, Nature chooses a signal distribution with knowledge of only the probability of allocation for each reported signal $q$ specified in $M$.

[17]Carrasco et al. (2018) analyzes the impact of the change in the seller's knowledge, measured by the change in the highest order of the moment that the seller knows, on the seller's guaranteed-profit from non-refundable offers.

[18]If the seller does not have any knowledge of the aggregate distribution of buyer valuations, i.e., even $\mu$ is unknown to the seller, then the seller cannot guarantee himself a positive profit.

# 3 Results

**The Best Guaranteed-Profit**

We begin this section by providing several important definitions necessary towards describing the paper's main findings. We then present the main results of the paper in Theorem 1, which identifies the *best guaranteed-profit* of the seller and a simple pricing policy, *the robust refund policy*, that allows him to achieve it. Finally, we explain why the robust refund policy provides some intuitive guarantee to the seller's profit before arguing that the robust refund policy provides the best guaranteed-profit of all possible mechanisms in $\mathcal{M}$. Furthermore, we will be stating our results in terms of the *normalized restocking cost* $\gamma \equiv \frac{c}{c+1}$ instead of $c$ for notational simplicity. We note that $\gamma$ is strictly increasing in $c$, $\lim_{c \to 0} \gamma = 0$ and $\lim_{c \to \infty} \gamma = 1$.

We first define a value $V_\gamma^* \in (0, 1)$ and a buyer's signal distribution $F_w$ characterized by $V_\gamma^*$ simultaneously. The function $F_w$ is illustrated in Figures 1(a) and 1(b) for low and high restocking cost $\gamma$ respectively. As we shall show, $V_\gamma^*$ and $F_w$ represent the seller's best guaranteed-profit and a worst-case distribution for the seller respectively. We define this worst-case distribution as:

$$F_w(q) \equiv \begin{cases} 0 & \text{if } q \in [0, V_\gamma^*), \\ 1 - \frac{V_\gamma^*}{q} & \text{if } q \in [V_\gamma^*, \gamma), \\ 1 - \min\{V_\gamma^*, \gamma\} & \text{if } q \in [\max\{V_\gamma^*, \gamma\}, 1), \\ 1 & \text{if } q = 1 \end{cases} \quad (1)$$

The distribution $F_w$ is a buyer's signal distribution if and only if $F_w \in \mathcal{F}$, i.e., $V_\gamma^*$ solves $\int_0^1 F_w(q) dq = 1 - \mu$. Since the equation $\int_0^1 F_w(q) dq = 1 - \mu$ has a unique solution $V_\gamma^*$ in $(0, 1)$, we can express $V_\gamma^*$ as follows:[19]

$$V_\gamma^* = \begin{cases} \frac{\mu - \gamma}{1 - \gamma} & \text{if } \gamma \leq \overline{\gamma} \equiv 1 - \sqrt{1 - \mu}, \\ \frac{\mu - V_\gamma^*}{1 - \gamma + \log \gamma - \log V_\gamma^*} = \frac{-\mu}{W_{-1}\left(-\frac{\mu}{\gamma} e^{\gamma - 2}\right)} & \text{if } \gamma > \overline{\gamma}. \end{cases} \quad (2)$$

We note that $V_\gamma^* \in (0, 1)$ is continuous and strictly decreasing in $\gamma$, with $V_1^* > 0$, $V_{\overline{\gamma}}^* = \overline{\gamma}$, and $V_0^* = \mu$. We provide a graphical presentation of $V_\gamma^*$ at different levels of $\gamma$ later in Figure 2.

We shall now name two pricing policies that are essential for our analysis. Both policies provide the seller with the means to which he can combat uncertainty.

**Definition 1** (Random Discounting). *Random discounting $P_{RD}$ is a pricing policy that consists of non-refundable offers $(p, 0)$ such that the price offered $p$ is log-uniformly distributed over the interval $[V_\gamma^*, \gamma]$.*[20]

---

[19]The function $W_{-1}(\cdot) \leq -1$ denotes the lower branch of the Lambert's W function, i.e., function $W_{-1}(x) = z$ is defined as the smaller of the two real solutions to the equation $ze^z = x$ for $z < 0$.

[20]The cumulative distribution of $p \in [V_\gamma^*, \gamma]$ is $\frac{\log p - \log V_\gamma^*}{\log \gamma - \log V_\gamma^*}$.

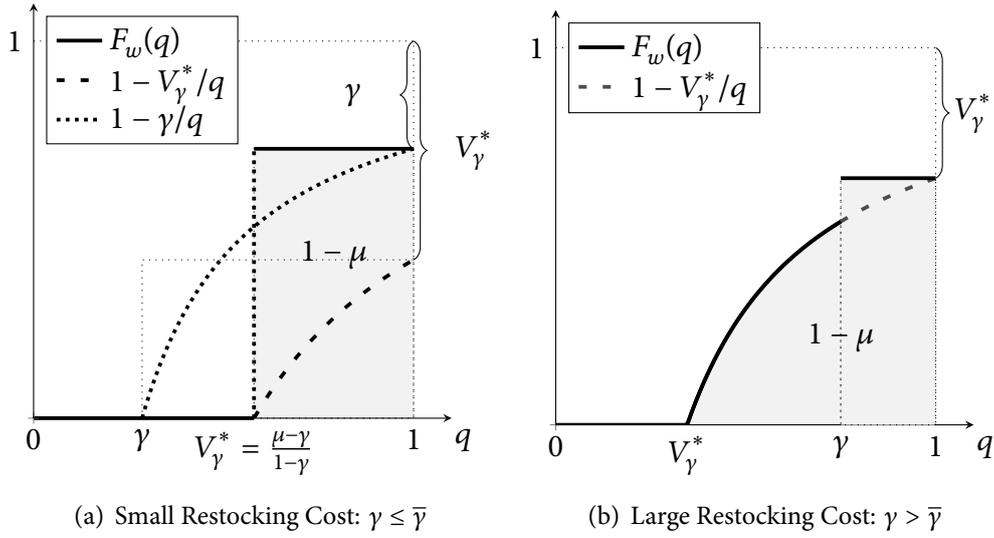

(a) Small Restocking Cost: $\gamma \leq \bar{\gamma}$

(b) Large Restocking Cost: $\gamma > \bar{\gamma}$

Figure 1: Best Guaranteed-Profit $V_\gamma^*$ and the Worst-Case Distribution $F_w$

Next, we consider refundable offers such that $\gamma \in (0, 1)$ is a *marginal signal*, i.e., the buyer is indifferent between buying and not buying when the signal is $\gamma$. Under these offers, the seller's profit from the buyer with marginal signal $\gamma$ is equal to zero regardless of her purchasing decision. Additionally, among such refundable offers is a fully refundable offer $(1, 1)$, which we call the *generous refund*.

**Definition 2** (Generous Refund). *The generous refund $P_{GR}$ is a refundable offer $(1, 1)$ such that the buyer buys if and only if $q \geq \gamma$.*

Among all possible mechanisms, a particular mixture of *random discounting* and *the generous refund* defined above allows the seller to achieve the best guaranteed-profit $V_\gamma^*$. We will define this randomization as follows:

**Definition 3** (Robust Refund Policy). *The robust refund policy $P_{RRP}$ is a pricing policy that induces the generous refund with probability $\beta_\gamma^*$ and random discounting with probability $1 - \beta_\gamma^*$, where*

$$\beta_\gamma^* = \begin{cases} 1 & \text{if } \gamma \leq \bar{\gamma}, \\ \frac{1-\gamma}{1-\gamma+\log \gamma - \log V_\gamma^*} & \text{if } \gamma > \bar{\gamma}. \end{cases} \quad (3)$$

Now that we have defined $V_\gamma^*$ and *the robust refund policy*, we are finally ready to formally state our first main result:

**Theorem 1.** *The best guaranteed-profit is $V_\gamma^*$, where $V_\gamma^*$ is defined in (2). The robust refund policy allows the seller to achieve the best guaranteed-profit. Furthermore, the best guaranteed-profit that the seller can achieve through random non-refundable offers is $V_1^*$, which is strictly smaller than $V_\gamma^*$ for all $\gamma < 1$.*

Theorem 1 makes three distinct claims. Firstly, the best guaranteed-profit by general mechanisms is $\sup_{M \in \mathcal{M}} \min_{F \in \mathcal{F}} V(M|F) = V_\gamma^*$. Secondly, the seller can achieve the best guaranteed-profit by the robust refund policy. As pricing policies are mechanisms, the best guaranteed-profit by pricing policies is bounded above by the best guaranteed-profit from general mechanisms, i.e.,

$$\min_{F \in \mathcal{F}} V(P_{RRP}|F) \leq \sup_{M \in \mathcal{M}} \min_{F \in \mathcal{F}} V(M|F). \tag{4}$$

Given that the seller can achieve the best guaranteed-profit by the robust refund policy, which is a pricing policy in $\mathcal{P}$, the inequality (4) holds with equality. Thirdly, without utilizing refunds, the seller cannot achieve the best guaranteed-profit. That is, the best guaranteed-profit by random non-refundable offers $V_1^*$ is strictly lower than $V_\gamma^*$.

## Robust Refund Policy: An Effective Hedge Against Uncertainty

We will now explain intuitively why the robust refund policy provides some guarantee on the seller's profit. In particular, we will first explain how the seller can use either of randomized prices or offering the generous refund to combat uncertainty. We will then explain how the robust refund policy is a pricing policy that, through combining these two tools, allows the seller to most effectively hedge against uncertainty.

**Randomizing Prices**

To understand how the seller hedges against uncertainty by randomizing prices, we start by considering the case where the seller makes a non-refundable offer $(p, 0)$. For a given price $p$, the buyer buys if and only if her signal $q$ is above the price $p$. Thus, a high price can result in a low profit due to a low probability of trade, especially when the buyer is relatively uninformed such that $q$ is likely to be much smaller than one. In contrast, a low price can result in a low profit due to a small profit margin conditional on a trade occurring, particularly when the buyer is relatively informed and thus $q$ is likely to be either close to 1 or 0.

We begin by noting that the *robust price*, i.e., the deterministic price that attains the seller's best guaranteed-profit by non-refundable offers, is $p_R^* = 1 - \sqrt{1 - \mu}$. Formally, deriving the guaranteed profit by a deterministic offer $(p, 0)$ is a canonical Bayesian persuasion problem in itself (Kamenica and Gentzkow (2011)). Nature chooses a signal distribution $F \in \mathcal{F}$ to minimize the seller's profit $v(p|q) = \mathbb{1}_{[q > p]} \times p$ with price $p$ and signal $q$. It is only when $p < \mu$ that a positive profit is guaranteed for the seller, with this worst-case profit being $p \times \frac{\mu - p}{1 - p} > 0$. Therefore, the seller can maximize his guaranteed-profit by setting $p = p_R^*$.

We then consider when the seller instead decides to randomize prices, i.e., setting different prices above or below the robust price with some probability. By offering prices below the robust price $p_R^*$ with positive probabilities, the seller can capture some gains from trade when the buyer's signal is likely to be moderate, i.e., $q$ is close to but lower than $p_R^*$. Thus, relative to the robust price, the seller's profit is higher when the buyer is relatively uninformed and is likely to have a moderate signal $q$. In addition, by offering prices above the robust price $p_R^*$ with positive probabilities, the seller earns a larger profit margin when the buyer's signal is favorable, i.e., $q > p_R^*$. Hence, relative to the robust price, we see that the seller's profit is higher when the buyer is relatively informed and is likely to have either a large $q$ or a small $q$.

More formally, we define the seller's best method of hedging against uncertainty via randomizing prices with non-refundable offers as that of *robust random pricing* $P_{RP}$, where the price offered is log-uniformly distributed on $[V_1^*, 1]$, with $V_1^*$ is defined in (2).[21] When the buyer's signal distribution is $F$, the seller's profit from robust random pricing is thus

$$V(P_{RP}|F) = \int_{V_1^*}^1 \frac{p[1 - F(p)]}{p(\log 1 - \log V_1^*)} dp = V_1^* + \frac{\int_0^{V_1^*} F(q)\,dq}{-\log V_1^*} \geq V_1^*. \quad (5)$$

From this, we present the following lemma:

**Lemma 1.** $V_1^*$ *is the best guaranteed-profit by randomizing prices.*

*Proof.* This is a special case of the exponential pricing identified in Du (2018). Let $G^{RS}$ be the distribution identified in Roesler and Szentes (2017), which is equivalent to $F_w$ for $\gamma = 1$:

$$G^{RS}(q) = \begin{cases} 0 & \text{if } q \in [0, V_1^*), \\ 1 - \frac{V_1^*}{q} & \text{if } q \in [V_1^*, 1) \\ 1 & \text{if } q = 1. \end{cases} \quad (6)$$

From this, it is straightforward to verify that the seller's profit by any non-refundable offer when the buyer's signal distribution is $G^{RS}$ is bounded from above by $V_1^*$, i.e., $V(P_{RP}|G^{RS}) \leq V_1^*$. Combining this with inequality (5) thus proves that $V_1^*$ is the best guaranteed-profit by randomizing prices. □

**Generous Refund**

The seller also can reduce the significance of the buyer's signal distribution on her purchasing decision by carefully designing a refund policy. Without loss of generality, we only discuss refundable offers $(p, r)$ such that $p \geq r$.[22] The buyer who faces refundable offer $(p, r)$ and receives a signal $q$ buys if and only if $q$ is above the marginal signal: $\tilde{q}(p, r) \equiv \frac{p-r}{1-r}$. We see that the marginal signal $\tilde{q}(p, r)$ is strictly less than price $p$ for all $p < 1$, and is strictly decreasing in $r$. That is, for a given price, the more generous the refund, the more likely the buyer purchases the product.

However, the seller needs to be wary of product returns. The buyer with signal $q$ who purchases the product returns it with probability $1-q$. For each product returned, the seller incurs the restocking cost $c = \frac{\gamma}{1-\gamma} > 0$. More specifically, if the seller makes a refundable offer $(p, r)$ and the buyer with signal $q$ buys the product, then his profit is:

$$\tilde{v}(q; p, r) \equiv p - (1-q)(c + r) = p - (1-q)\left(\frac{\gamma}{1-\gamma} + r\right). \quad (7)$$

If the refund is too generous, the buyer buys even when the value of $q$ is small, causing the seller to suffer a significant loss, i.e., $\tilde{v}(q; p, r)$ is negative for $q$ close to but above the marginal signal. In contrast, if the refund offered is too low, then the seller fails to induce the buyer with a moderate signal to buy, i.e., even though $\tilde{v}(q; p, r)$ is positive for $q$ close to but below the marginal signal, the buyer with such a $q$ would not buy.

---

[21]That is, $V_1^*$ is the unique solution to $\int_V^1 \left(1 - \frac{V}{q}\right) dq = 1 - \mu$. The CDF that represents robust random pricing takes 0 for $q \in [0, V_1^*]$ and $\frac{\log q - \log V_1^*}{-\log V_1^*}$ for $q \in [V_1^*, 1]$.

[22]Notice that if $p < r$, then the buyer buys irrespective of the value of the signal. Raising price by a small amount therefore strictly increases the profit.

However, it is important to note that the seller does not know the distribution of $q$. Thus, the seller can best hedge against this uncertainty by designing a refund policy such that the buyer buys if and only if she brings in a positive profit conditional on the purchase being made. The seller does this by assigning an amount to $r(p)$ so that $\tilde{v}(q; p, r(p)) \geq 0$ if and only if $q \geq \tilde{q}(p, r(p))$. He can achieve this by setting the refund to $r(p) = \frac{p-\gamma}{1-\gamma}$ so that the marginal signal is $\tilde{q}(p, r(p)) = \gamma$. Furthermore, the seller can strictly increase his profit from a buyer with signal $q > \gamma$, i.e., $\tilde{v}(q; p, r(p)) = \frac{q-\gamma}{1-\gamma}p$, by increasing both the price $p$ and refund offered $r(p)$ to $(1, r(1)) = (1, 1)$. It is this very argument that defines the generous refund provided in Definition 2.[23]

**Robust Refund Policy**

Having identified two simple pricing policies that the seller can use to tackle uncertainty, i.e., randomizing prices and the generous refund; we now discuss why the robust refund policy defined in Definition 3 allows the seller to better hedge against uncertainty in general.

The generous refund policy, by design, screens the buyer based on whether her signal is above $\gamma$. Therefore, it is effective when the buyer's signal is likely to be favorable, i.e., when the buyer is relatively informed, but not so when the buyer is relatively uninformed. More formally, when the buyer's signal distribution is $F$, the seller's profit from using the generous refund policy is:

$$V(P_{GR}|F) = \int_\gamma^1 \frac{q-\gamma}{1-\gamma} dF(q) = 1 - \frac{\int_\gamma^1 F(q)\, dq}{1-\gamma} \geq 0. \qquad (8)$$

We see that the profit from using the generous refund policy is decreasing in $\int_\gamma^1 F(q)\, dq$. Observe that the term $\int_\gamma^1 F(q)\, dq$ being small implies that $F$ is flat and that the likelihood of the signal $q$ being above $\gamma$ is high, i.e., the buyer is relatively informed. Thus, when the buyer is relatively informed, the generous refund brings in a significant profit. In contrast, if $F$ is steep and the signal is concentrated around $\mu$ so that $\int_\gamma^1 F(q)\, dq$ is large, i.e., when the buyer is relatively uninformed, then the profit from using the generous refund is small.

To hedge against the possibility of the latter, the seller can randomize prices in a way that is more likely to generate a moderate price than that under robust random pricing. More specifically, suppose that $\gamma \geq \bar{\gamma}$ so that $V_\gamma^* \leq \gamma$. With random discounting $P_{RD}$, i.e., the log-uniform randomization of prices over $[V_\gamma^*, \gamma]$ for $V_\gamma^* \in (V_1^*, \gamma)$ as defined in Definition 1, the profit of the seller who faces the buyer with signal distribution $F$ is

$$V(P_{RD}|F) = \int_{V_\gamma^*}^\gamma \frac{p[1-F(p)]}{p\left(\log\gamma - \log V_\gamma^*\right)} dp = \frac{\gamma - V_\gamma^* - \int_{V_\gamma^*}^\gamma F(q)\, dq}{\log\gamma - \log V_\gamma^*}. \qquad (9)$$

---

[23]If $p < 1$, then $q = \gamma$ is the unique marginal signal under $(p, r(p))$. However, if $p = 1$, then $r(1) = 1$ and hence all signals are marginal signals. Therefore, one may interpret the generous refund as the limit of $(p, r(p))$ where $p$ is taken to 1; specified as the buyer's tie-breaking rule at $(1, 1)$.

Comparing the seller's profit from random discounting $V(P_{RD}|F)$ with the one from robust random pricing $V(P_{RP}|F)$ derived in (5), we find that the former is likely to bring in a larger profit when the buyer is relatively uninformed such that $F$ is steep.

Given that the generous refund works well with a relatively informed buyer whereas random discounting performs better with a relatively uninformed buyer, we expect that a combination of both of these policies we call the robust refund policy will be able to allow the seller to hedge against uncertainty to an even greater degree.

Observe that the buyer's signal distribution $F$ only affects the seller's profit under the generous refund through $\int_\gamma^1 F(q)\,dq$; and the profit under random discounting through $\int_{V_\gamma^*}^\gamma F(q)\,dq$. Thus, it is possible for the seller to reduce the impact of uncertainty that $F$ has on his profit if he can randomize over both pricing policies such that his profit only depends on the sum of both integrals, i.e., $\int_{V_\gamma^*}^1 F(q)dq$, rather than each term independently.

The seller can do so by choosing the value of $\beta$ such that the coefficient on $\int_\gamma^1 F(q)\,dq$, i.e., $\frac{\beta}{1-\gamma}$, and the coefficient on $\int_{V_\gamma^*}^\gamma F(q)\,dq$, i.e., $\frac{(1-\beta)(\gamma-V_\gamma^*)}{\log\gamma-\log V_\gamma^*}$ are equal.[24] It is this very argument that defines the weight $\beta$ in Definition 3. Then, we may write the seller's profit from the robust refund policy when the buyer's signal distribution is $F$ as:

$$V(P_{RRP}|F) = \beta_\gamma^* V(P_{GR}|F) + \left(1-\beta_\gamma^*\right)V(P_{RD}|F)$$

$$= \begin{cases} V_\gamma^* + \frac{\int_0^\gamma F(q)dq}{1-\gamma} & \text{if } \gamma \leq \bar{\gamma}, \\ V_\gamma^* + \frac{\int_0^{V_\gamma^*} F(q)dq}{1-\gamma+\log\gamma-\log V_\gamma^*} & \text{if } \gamma > \bar{\gamma}. \end{cases} \quad (10)$$

First, we see that the seller's guaranteed-profit through using the robust refund policy is $V_\gamma^*$ because $V(P_{RRP}|F_w) = V_\gamma^*$. We also note that $V(P_{RRP}|F) > V(P_{RP}|F)$ for all $F$ and $\gamma < 1$, i.e., for any given buyer's signal distribution, the profit from the robust refund policy is always strictly higher than the profit under robust random pricing. This is because $V(P_{RRP}|F)$ is strictly decreasing in $\gamma$ when $\gamma > \bar{\gamma}$ and approaches $V(P_{RP}|F)$ as $\gamma \to 1$.

To further illustrate this, we provide a comparison of the guaranteed-profit by various pricing policies as the functions of the normalized restocking cost $\gamma$ in Figure 2. In particular, we see that $V_\gamma^*$ is strictly decreasing in $\gamma$, and hence $V_\gamma^* > V_1^*$ for all $\gamma < 1$. Thus, the guaranteed-profit by the robust refund policy is strictly higher than the guaranteed-profit by robust random pricing. This leads us to provide the following lemma:

**Lemma 2.** *For any buyer's signal distribution $F \in \mathcal{F}$, the seller achieves a strictly greater profit under the robust refund policy than that under robust random pricing. Consequently, the seller's guaranteed-profit under the robust refund policy is strictly greater than that under robust random pricing, and strictly decreasing in the normalized restocking cost $\gamma$.*

## Worst-Case Distribution

We have argued that the robust refund policy $P_{RRP}$ offers an intuitive guarantee on profit, i.e., $\min_{F \in \mathcal{F}} V(P_{RRP}|F) = V_\gamma^*$; and it is strictly higher than the best guaranteed-profit by random pricing, i.e., $V_\gamma^* > V_1^*$. We will now proceed to show that the robust refund policy provides the best guaranteed-profit among all possible mechanisms, i.e., $\min_{F \in \mathcal{F}} V(M|F) \leq V_\gamma^*$ for all $M \in \mathcal{M}$. To do this, we will first demonstrate that the robust refund policy provides the best guaranteed-profit among all possible pricing policies. We then show that pricing policies can provide the best guaranteed-profit among all possible mechanisms. These are represented by having the following inequalities hold with equality:

$$\sup_{P \in \mathcal{P}} \min_{F \in \mathcal{F}} V(P|F) \geq \min_{F \in \mathcal{F}} V(P_{RRP}|F) = V_\gamma^*; \tag{11}$$

$$\sup_{M \in \mathcal{M}} \min_{F \in \mathcal{F}} V(M|F) \geq \sup_{P \in \mathcal{P}} \min_{F \in \mathcal{F}} V(P|F). \tag{12}$$

---

[24]If the seller induces the generous refund with probability $\beta$ and random discounting with probability $1-\beta$, then his profit is $\beta + \frac{(1-\beta)(\gamma-V_\gamma^*)}{\log\gamma-\log V_\gamma^*} - \left(\frac{\beta}{1-\gamma}\int_\gamma^1 F(q)\,dq + \frac{(1-\beta)}{\log\gamma-\log V_\gamma^*}\int_{V_\gamma^*}^\gamma F(q)\,dq\right)$.

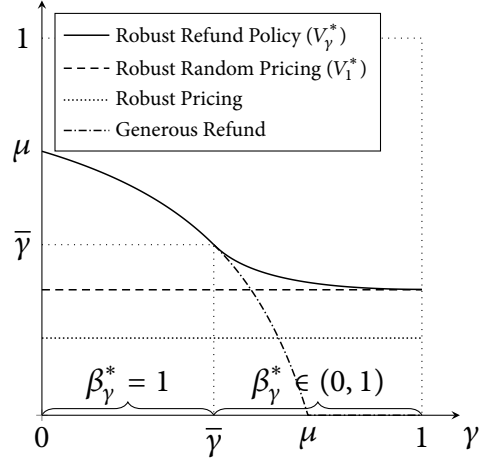

Figure 2: Guaranteed-Profit under Various Pricing Policies

We first consider when $\gamma = 0$, i.e., when the seller can offer refunds costlessly. In this case, the seller's guaranteed-profit from the generous refund is $\mu = \lim_{\gamma \to 0} V_\gamma^* = V_0^*$. As $\mu$ is the upper bound for gains from trade, this implies that the seller can completely hedge against uncertainty through using the robust refund policy, which, in this case, coincides with the generous refund. Thus the robust refund policy provides the best guaranteed-profit when $\gamma = 0$.

We now consider when $\gamma > 0$. Observe that the robust refund policy provides the best guaranteed-profit among all possible pricing policies if and only if, for $F_w$ defined by (1), all offers in the support of the robust refund policy are profit maximizing offers and generate a profit of $V_\gamma^*$. As both refundable and non-refundable offers comprise the set of $\mathcal{P}$, we will discuss these cases separately, showing that the seller's profit from either type of offer under $F_w$ is bounded above by $V_\gamma^*$.

First, recall that by (8), the seller's profit under the generous refund policy, which induces a marginal signal $\gamma$, is $V(P_{GR}|F_w) = V_\gamma^*$. This policy is always in the support of the robust refund policy. Among all refundable offers that induce a marginal signal of $\gamma$, the generous refund provides the highest profit. Thus, the seller's profit from a refundable offer is bounded above by $V_\gamma^*$. Additionally, with refundable offers, the sellers wishes for the buyer to purchase if and only if she brings in a positive profit conditional on purchasing the product. For otherwise, the seller's profit would be bounded above $V_\gamma^*$. Thus, we see that among all refundable offers, the generous refund brings the seller the highest profit.

Next, as discussed in Roesler and Szentes (2017), when the buyer's signal distribution is $F$, the supremum of the seller's profit from a non-refundable offer is $V$ if $\lim_{\tilde{p} \to p^-} F(\tilde{p}) = 1 - V/p$ and $F(q) \geq 1 - V/q$ for all $q \in [0, 1]$. Therefore, when the buyer's signal distribution is $F_w$, the seller's profit from any non-refundable offer is bounded above by $V_\gamma^*$.[25] Furthermore, a non-refundable offer is in the support of the robust refund policy if and only if $\gamma > \overline{\gamma}$, and if so, all non-refundable offers in the support of the robust refund policy, i.e., $(p, 0)$ such that $p \in [V_\gamma^*, \gamma]$, generate a profit of $V_\gamma^*$. We therefore conclude that among all pricing policies, the robust refund policy provides the seller with the highest profit, i.e., $\sup_{P \in \mathcal{P}} \min_{F \in \mathcal{F}} V(P|F) = V_\gamma^* \leq V(P_{RPP}|F)$ for all $F \in \mathcal{F}$.

We now show that the robust refund policy also provides the best guaranteed-profit among all mechanisms in general. Recall that the robust refund policy essentially screens the buyer based on his realized signal $q$ without utilizing any information about the signal's underlying distribution.[26] Given this, a seller who can obtain any (albeit impartial) information about the buyer's signal distribution can use it to his advantage to screen the buyer based on her realized signal and earn a higher profit. A priori, nothing seems to guarantee that a worst-case distribution for the seller who can only utilize pricing policies, i.e., mechanisms in $\mathcal{P} \subset \mathcal{M}$, is a worst-case distribution for the seller who can utilize any mechanism in $\mathcal{M}$. Even if it is, it could be possible for a non-pricing policy $M \in \mathcal{M} \setminus \mathcal{P}$ to provide the best guaranteed-profit. To address this, we will show that $F_w$ is a *worst-case distribution* among *all mechanisms*, i.e., $F_w \in \arg\min_{F \in \mathcal{F}} \max_{M \in \mathcal{M}} V(M|F)$, and that the upper bound of the profit against $F_w$ is $V_\gamma^*$, i.e., $\sup_{M \in \mathcal{M}} V(M|F_w) = V_\gamma^*$.

To provide some context with regards to what we are about to show, suppose hypothetically that the buyer's signal distribution is commonly known to be some $F \in \mathcal{F}$. It is well known in mechanism design literature that if the seller and the buyer can only interact once and if the distribution is regular, a simple non-refundable offer will be a profit maximizing mechanism.[27] Analogously, we demonstrate that in our case, where the seller and the buyer can interact more than once through product return and if buyer's signal distribution of $F_w$ is commonly known, the seller will be able to maximize his profit through a refundable or non-refundable offer. Furthermore, we have already established that against $F_w$, the seller's profit from any pricing policy is bounded from above by $V_\gamma^*$, the profit that is provided by the robust refund policy. We can therefore conclude that the robust refund policy provides the best guaranteed-profit not only among all pricing policies, but also among all possible mechanisms. In this sense, the distribution $F_w$ is a distribution that leaves no room for the seller to take advantage of other more complicated mechanism.

We will now formalize our argument. Suppose that the buyer's signal distribution is commonly known. A profit maximizing mechanism can be found from the following class of mechanisms that utilize refunds:

---

[25] Observe that the solid black curve ($F_w$) is weakly above the loosely dotted gray curve $(1 - V_\gamma^*/q)$ for all $q$, and they coincide on $[\gamma, V_\gamma^*]$ in Figures 1(a) and 1(b).

[26] Observe that the probability that the buyer receives the product (without the option to return) is strictly increasing on $[V_\gamma^*, \gamma]$.

[27] By interacting once, we mean that the interaction between the buyer and the seller ends with the purchase (or equivalently in our model, when restocking costs are infinitely large).

**Definition 4** (Direct Mechanism with Refunds). *An individually-rational and incentive-compatible direct mechanism*

$$M_D \equiv \{p(q), \{\alpha_0(q), \alpha_r(q)\}\}_{q \in [0,1]}$$

*is a direct mechanism with refunds if, for each buyer's report $q \in [0, 1]$, the mechanism specifies (i) $p(q)$: the transfer from the buyer to the seller; (ii) $\alpha_0(q) \in [0, 1]$: the probability that the buyer receives the product without an option to return; and (iii) $\alpha_r(q) \in [0, 1 - \alpha_0(q)]$: the probability that the buyer receives the product with an option to return with refund $r = 1$.*

**Lemma 3.** *When the buyer's signal distribution is commonly known to be F, there exists a direct mechanism with refunds $M_D$ that maximizes the seller's profit, i.e., $V(M_D|F) = \max_{M \in \mathcal{M}} V(M|F)$.*

*Proof.* By the revelation principle, for any indirect mechanism $M \in \mathcal{M}$, there exists an outcome-equivalent dynamic direct mechanism that is individually-rational and incentive-compatible. The formal proof in the appendix shows that such a dynamic direct mechanism is implementable through a direct mechanism with refunds. □

Lemma 3 allows us to apply the standard argument to show that the generous refund is a profit maximizing mechanism. More formally, we can represent the seller's profit from a direct mechanism with refunds $M_D$ when the buyer's realized signal is $q$, which we denote by $v(q|M_D)$, as follows:

**Lemma 4.** *Suppose that the buyer's signal distribution is commonly known to be F. If $M_D = \{p(q), \{\alpha_0(q), \alpha_r(q)\}\}_{q \in [0,1]}$ is a profit-maximizing mechanism, then*

$$\alpha_0(q) \text{ is (weakly) increasing in } q \tag{13}$$

*and*

$$v(q|M_D) \equiv \begin{cases} q\alpha_0(q) - \int_0^q \alpha_0(\tilde{q}) d\tilde{q} & \text{if } q < \gamma, \\ q\alpha_0(q) + \frac{q-\gamma}{1-\gamma}(1 - \alpha_0(q)) - \int_0^q \alpha_0(\tilde{q}) d\tilde{q} & \text{if } q \geq \gamma. \end{cases} \tag{14}$$

*Proof.* In the Appendix. □

We can therefore further simplify the seller's problem to the one in which he chooses a (weakly) increasing function $\alpha_0(\cdot)$, instead of a triplet of functions $M_D$, to maximize $V(\widetilde{M}|F)$. We then find that when the buyer's signal distribution is $F_w$, the function $\alpha_0(q)$, i.e., the probability that the buyer with posterior $q$ receives the product without an option to return, is indeterminate; and all offers in the support of the robust refund policy turn out to be profit maximizing mechanisms.

**Lemma 5.** *Suppose that the buyer's signal distribution is commonly known to be $F_w$ as defined in (1). The seller's profit from a profit-maximizing mechanism, i.e., $\max_{\widetilde{M} \in \mathcal{M}} V(\widetilde{M}|F_w)$, is $V_\gamma^*$. Thus, $F_w$ is a worst-case distribution among all mechanisms, and the robust refund policy provides the best guaranteed-profit $V_\gamma^*$.*

*Proof.* In the Appendix. □

We emphasize that our goal here is not to identify optimal mechanisms for all possible buyer's signal distributions. In particular, Lemma 5 only informs us that the robust refund policy is a profit-maximizing mechanism when the buyer's signal distribution is $F_w$; for an arbitrary buyer's signal distribution, the lemma is silent about the differences between the robust refund policy and optimal mechanisms, as well as any differences in the profits that they generate for the seller, which can be as large as $\mu - V_\gamma^*$, when we consider an arbitrary buyer's signal distribution.

Instead, our primary interest is to identify a mechanism that works well under all possible buyer signal distributions, including the ones that do not satisfy a set of certain regularity conditions that are often assumed in the literature.[28] As formally stated in Theorem 1, we show that the robust refund policy provides the best guaranteed-profit among all possible mechanisms $V_\gamma^*$, including that of both pricing and non-pricing mechanisms, and other potentially more complicated mechanisms. Moreover, the seller's best guaranteed-profit under the robust refund policy is strictly higher than the best guaranteed-profit under randomized prices, including that of robust random pricing.

To conclude this section, we note that our results are closely related to Roesler and Szentes (2017) and Du (2018). In particular, if (and only if) the restocking cost is infinitely high so that $\gamma = 1$, our worst-case distribution $F_w$, the robust refund policy, and the best guaranteed-profit, respectively correspond to the worst-case distribution, the robust mechanism, and the best guaranteed-profit identified in Du (2018). In contrast to these articles, we take into account the possibility of interactions after the initial transaction between the buyer and the seller, which itself is a part of the seller's design problem. Therefore, there exists an interplay between the buyer's signal distribution, the buyer's purchasing decision, and the mechanisms the seller can offer in our model that is absent in previous works. This interplay is potentially non-trivial, as it creates a room for the seller to take advantage of sequential screening mechanisms to discriminate the buyer with different posterior beliefs. Our results show that the seller, who is interested in maximizing his guaranteed-profit, chooses to continue interacting with the buyer after the initial transaction through offering the generous refund with a strictly positive probability.

## 4 Buyer-Optimal Information Structure

The analysis in the previous section identified a worst-case distribution for the seller. We now provide an answer as to when, if ever, the seller should be concerned about such a worst-case. We do so by showing that the worst-case distribution arises as an equilibrium of the game described below in which the buyer strategically decides what to learn. The game proceeds as follows:

1. The buyer chooses and commits to a signal distribution $F \in \mathcal{F}$. The choice of signal distribution is costless.

2. The seller observes $F$, and commits to a mechanism $M \in \mathcal{M}$.

---

[28] See Inderst and Tirosh (2015); Ely et al. (2017); Courty and Li (2000); Eső and Szentes (2007); Pavan et al. (2014) for an analysis of related settings, where the buyer's signal distribution is commonly known and satisfies various regularity conditions.

3. The buyer learns of her signal $q \sim F$, and the buyer and the seller interact according to the rules of the mechanism $M$.

We note that the critical assumption we impose is that the buyer commits to a signal distribution at the first stage of the game. In other words, the buyer commits to not learning any information that is not contained in the signal generated by distribution $F$.[29] Note that if the buyer lacks such a commitment power, then she would choose to acquire a fully informative signal. Knowing that the buyer chooses a fully informative signal, the seller would then always set the price to 1 and thus capture the full surplus of the buyer.

We consider the equilibrium that maximizes the buyer's equilibrium payoff and call her equilibrium strategy *the buyer-optimal information structure*. If $\gamma = 1$, i.e., the restocking cost is infinitely high, then this game coincides with the game analyzed in Roesler and Szentes (2017). As analyzed in Du (2018), in the absence of after-purchase buyer-seller interactions or when the restocking cost is infinitely high, i.e., $\gamma = 1$, the buyer-optimal information structure is known to coincide with the seller's worst-case distribution.

Let $U(M|F)$ denote buyer's expected payoff when she chooses the signal distribution $F$ and the seller responds by choosing the mechanism $M$. Then the buyer's problem can be described as

$$\max_{F \in \mathcal{F}, M^* \in \mathcal{M}} U(M^*|F) \text{ subject to } M^* \in \arg\max_{M \in \mathcal{M}} V(M|F) \qquad (15)$$

That is, the buyer essentially chooses a signal distribution and "recommends" a mechanism to the seller that the seller finds optimal to follow given the buyer's signal distribution.

Lemma 3 allows us to focus on direct mechanisms with refunds instead of all possible mechanisms. Yet, at a glance, this problem forces us to identify all profit maximizing direct mechanisms with refunds for each possible $F \in \mathcal{F}$. Nevertheless, as an immediate corollary of the analysis conducted in the prior section, we can show that the buyer-optimal information structure coincides with the seller's worst-case distribution $F_w$ defined in (1).

Formally, the analysis in the previous section shows that (i) regardless of the buyer's strategy, the seller's profit is bounded from below by his best guaranteed-profit, i.e., $V_\gamma^*$; and (ii) if the buyer chooses $F_w$, then making the non-refundable offer $(V_\gamma^*, 0)$ is one of the seller's best responses. As the gain from trade is $\mu$, the buyer's equilibrium payoff cannot exceed $\mu - V_\gamma^*$. However, if the buyer chooses $F_w$ and the seller best responds by offering $(V_\gamma^*, 0)$, then the trade occurs with probability one, and the buyer's payoff would be $\mu - V_\gamma^*$, i.e., the upper bound. This establishes that $F_w$ is a buyer-optimal information structure, allowing us to state the following theorem:

**Theorem 2.** *In the game in which the buyer chooses and commits to a signal distribution F; and the seller best responds by a mechanism, $F_w$ is a buyer-optimal information structure.*

In this sense, if the seller cannot exclude the possibility that the buyer believes that the seller optimally responds to the choice of buyer's signal distribution, then the seller has a good reason to be concerned of the possibility of facing the worst-case distribution.

---

[29] A possible interpretation of this model is as follows: the buyer ex-ante commits to delegating the information gathering to a third-party, such as an agent or an algorithm, knowing that the seller best responds to the signal distribution by choosing a pricing policy.

# 5   Discussion

We analyze the robust pricing problem of an experience good seller who faces uncertainty about a potential buyer's prior information and learning. Our results demonstrate that a simple mechanism that utilizes both randomized prices and the generous refund provides the best guaranteed-profit. Here, we provide a discussion of several of the underlying assumptions we make and areas for potential future research.

To make further predictions about how the seller's use of the robust refund policy may change in different environments, it is helpful to discuss how our results will change if the value of the product to the buyer conditional on a good fit is $\tilde{v} > 0$ instead of 1. In our analysis, the normalized restocking cost $\gamma$ captures both the restocking cost $c$ as well as the marginal signal under the generous refund. Noting this, we define the normalized restocking cost as $\tilde{\gamma} \equiv \frac{c\tilde{v}}{\tilde{v}+c} \in (0, \tilde{v})$. Then, the generous refund will be $(p, r) = (\tilde{v}, \tilde{v})$ with marginal signal $\tilde{\gamma}/\tilde{v} \in (0, 1)$, and random discounting will be a pricing policy that consists of non-refundable offers $(p, 0)$ such that $p/\tilde{v}$ is log-uniformly distributed over the interval $\left[V^*_{\tilde{\gamma}/\tilde{v}}, \tilde{\gamma}/\tilde{v}\right]$. Additionally, the robust refund policy induces the generous refund with probability $\beta^*_{\tilde{\gamma}/\tilde{v}}$ and random discounting with probability $1 - \beta^*_{\tilde{\gamma}/\tilde{v}}$. The seller's best guaranteed-profit is $\tilde{v} \times V^*_{\tilde{\gamma}/\tilde{v}}$. The marginal signal for $\tilde{v}$, i.e., $\tilde{\gamma}/\tilde{v}$ is decreasing in $\tilde{v}$ and increasing in the restocking cost $c$, and $\beta^*_{\tilde{\gamma}/\tilde{v}}$ is decreasing in $\tilde{\gamma}/\tilde{v}$. Therefore, both an increase in the match value $\tilde{v}$ and a decrease in the restocking cost $c$ lead to an increase in the probability that the buyer is offered the generous refund.

For instance, consider a fashion retailer. At the beginning of the season, the buyer may have a higher match value as the product is "fashionable". In other words, $\tilde{v}$ goes down over time during the season. For the same reason, the retailer may also find that returned products have a larger salvage value at the beginning of the season than at the end of the season, particularly as they will then have more time to resell the product before the product goes "out of fashion." We may interpret this as the (non-normalized) restocking cost $c$ going up over time. Our model would thus predict that the retailer wishes to make full-price fully-refundable offers (generous refund policy) early in the season, and make "all sales final" offers near the end of the season. Such policies are a common practice by many retailers. For example, GAP, Guess, Zumiez, and Banana Republic offer a full refund for items returned in their original condition, but make an exception for final sale items, which cannot be returned or exchanged.

Our model hinges on the restrictive assumption of the buyer's valuation being binary, which significantly simplifies the analysis conducted. This assumption is reasonable if the buyer's primary concerns are of the exact product match: e.g., shoes and clothes either do or don't fit, a gadget is either compatible with the buyer's use or not, and a business traveler either needs to be in a particular location on a specific date or not. We note that the implications of this assumption are twofold. Firstly, there exists a one-to-one mapping between the signal that the buyer receives and her expected willingness to pay for the product (in the absence of a refund policy). Secondly, the buyer's return decision is independent of the refund amount offered.

Those two properties fail to hold if more than two levels of product fit are present: e.g., when clothes do not fit at all, fit somewhat well or fit perfectly. Thus, the assumption of binary buyer valuation, whereas restrictive, is important for the tractability of the model. More precisely, in the absence of a refund policy, the buyer buys only if her signal $q$, i.e., a posterior over possible levels of product fit $v$, satisfies $\mathbb{E}_q[v] \geq p$. That is, the seller's profit depends on the buyer's signal $q$ only through the value of expected willingness to pay it induces, i.e., $\mathbb{E}_q[v]$. In this sense, if the seller cannot utilize a refund policy, then we still can represent the seller's uncertainty as the uncertainty over distributions of the buyer's expected willingness to pay, which is a one-dimensional random variable.[30] In the presence of a refund policy $(p, r)$, however, the buyer with signal $q$ buys only if the right-tail of $q$ is sufficiently fat, i.e., $\Pr(v > r|q) \mathbb{E}_q[v|v > r] + \Pr(v \leq r|q) r \geq p$. Therefore, two signals with an identical willingness to pay can result in different outcomes for the seller. Consequently, we are no longer able to capture the seller's uncertainty in terms of the uncertainty over distributions of the buyer's expected willingness to pay, rendering the generalization that we make in this direction not as straightforward.

Having said that, we conjecture that the relaxation of the binary buyer valuation assumption yields qualitatively similar results: for any given restocking cost, there must exist a generous refund policy (or randomization over refund policies) that guarantees the seller a non-negative ex-post (expected) profit over all possible buyer's signal distributions. Hence, combining such a refund policy with random pricing is still likely to guarantee a higher profit than simply using randomized prices alone.

We also abstract away the possibility of buyer heterogeneity in our model. In reality, the seller may face two or more groups of consumers. If the seller could differentiate between groups, then our analysis can be applied to each group separately. We illustrate this with an example. Consider an airline that faces both corporate clients and private consumers. Suppose that corporate clients have either higher match valuation or lower transaction costs for returns. As discussed above, our results then imply that corporate clients will receive the generous refund offers more frequently, whereas private customers will receive non-refundable discounted offers more often. A similar third-degree price-discrimination argument applies in other situations, such as geographic differentiation.

In other situations, it may be more realistic to assume that the seller cannot distinguish different groups of buyers and has only a prior belief about which group a particular consumer belongs to. Our model will naturally extend to this setting if the seller's partial knowledge is about the first moment of the population of each group. Should we be able to identify a "worst-case scenario" for the seller, we conjecture that an optimal mechanism can be shown to be a menu that consists of the generous refund and random discounting. However, a major roadblock is the identification of this worst-case scenario, which will be a profile of buyer signal distributions for each group. The profile that consists of the worst-case distributions for homogeneous buyers, one of which we identify and analyze in this paper, is unlikely to be the worst-case scenario within this context.

Another aspect that we do not address is the seller's learning of buyer demand through pricing. We believe that the insight we provide, that a well-designed refund policy limits the significance of the buyer's learning on the seller's profit, should apply even within the context of a dynamic environment. Additionally, a carefully designed dynamic pricing policy with refunds, and the buyer's resulting purchasing and return decision can be used by the seller to learn about what and how the buyer learns about the product's fit. When faced with repeated purchases or buyers arriving sequentially over time, this information can also be used by the seller to help improve his future pricing decisions. Investigating how the seller's learning motive would shape intertemporal pricing with refunds would be an interesting avenue for future research.

---

[30] More precisely, the distributions (over signals) that minimize the seller's profit from a deterministic non-refundable offer and the seller's worst-case distribution over signals when the seller cannot utilize a refund policy can both be characterized in terms of distributions over the buyer's expected willingness to pay.

Finally, we note that our article's findings highlight the importance of disclosure rules with regards to platform design problems. A third-party platform, such as eBay or Airbnb, can choose which information it reveals to the buyer about the product offered: for example, details of the product description, photos and videos, and ratings and reviews of previous consumers. The seller then chooses a pricing and return policy, and the buyer sequentially makes a purchasing decision, potentially making a return decision upon learning more about the product after purchase. Our buyer-optimal learning model shows that the platform can maximize the volume of trade by disclosing the information about product match in a way that maximizes the buyer's surplus. A valuable question to explore in future research would be what information the platform chooses to disclose when its objective is not to maximize the volume of trade, e.g., maximize the total revenue from the sales.

# Appendix: Proofs

**Proof of Lemma 3**

A dynamic direct mechanism $M_2 = \left\{\alpha_q, p_q, \left\{\left(\kappa_q^v, \tau_q^v\right)\right\}_{v \in \{0,1\}}\right\}_{q \in [0,1]}$ is a two-step mechanism if it specifies, for each reported signal $q \in [0, 1]$, (i) $\alpha_q$: the probability that the buyer receives the product; (ii) $p_q$: the transfer from the buyer to the seller; and (iii) $\left\{\kappa_q^v, \tau_q^v\right\}_{v \in \{0,1\}}$: the direct mechanism that specifies, for each buyer's reported realized valuation $v \in \{0, 1\}$, the probability that the buyer keeps the product $\kappa_q^v$ and the transfer from the seller to the buyer $\tau_q^v$ such that $\kappa_q^v v + \tau_q^v \geq \max\left\{v, \kappa_q^{v'} v + \tau_q^{v'}\right\}$ for $v \in \{0, 1\}$ and $v' \neq v$.

If the buyer with signal $q$ reports $q'$ under the two-step mechanism $M_2$, then her payoff is

$$u\left(q, q' | M_2\right) \equiv \alpha_{q'} q \left(\kappa_{q'}^1 + \tau_{q'}^1\right) + \alpha_{q'} (1 - q) \tau_{q'}^0 - p_{q'}.$$

Furthermore, a two-step mechanism $M_2$ is an IR-IC two-step mechanism if $u(q, q | M_2) \geq 0$ for all $q$ and $u(q, q | M_2) \geq u\left(q, q' | M_2\right)$ for all $q$ and $q'$. The seller's profit from the buyer with signal $q$ under an IR-IC two-step mechanism $M_2$ is

$$v(q|M_2) \equiv p_q - \alpha_q q\left(\left(1 - \kappa_q^1\right)c + \tau_q^1\right) - \alpha_q \times (1-q)\left(\left(1 - \kappa_q^0\right)c + \tau_q^0\right)$$
$$= -u(q, q|M_2) - \alpha_q q\left(\left(1 - \kappa_q^1\right)c - \kappa_q^1\right) - \alpha_q \times (1-q)\left(1 - \kappa_q^0\right)c.$$

If $M$ is a profit-maximizing mechanism, then by the revelation principle, there exists an outcome-equivalent IR-IC two-step mechanism $\widetilde{M_2} = \left\{\widetilde{\alpha}_q, \widetilde{p}_q, \left\{\left(\widetilde{\kappa}_q^v, \widetilde{\tau}_q^v\right)\right\}_{v\in\{0,1\}}\right\}_{q\in[0,1]}$. Define (i) $\alpha_q = \widetilde{\alpha}_q$; (ii) $p_q = \widetilde{p}_q + \widetilde{\alpha}_q\left(1 - \left(\widetilde{\kappa}_q^1 + \widetilde{\tau}_q^1\right)\right)$; (iii) $\left(\kappa_q^0, \tau_q^0\right) = \left(\widetilde{\kappa}_q^1 + \widetilde{\tau}_q^1 - \widetilde{\tau}_q^0, 1 + \widetilde{\tau}_q^0 - \left(\widetilde{\kappa}_q^1 + \widetilde{\tau}_q^1\right)\right)$; and (iv) $\left(\kappa_q^1, \tau_q^1\right) = (1, 0)$. We first show that $M_2 \equiv \left\{\alpha_q, p_q, \left\{\left(\kappa_q^v, \tau_q^v\right)\right\}_{v\in\{0,1\}}\right\}_{q\in[0,1]}$ is an IR-IC two-step mechanism, and outcome-equivalent to $\widetilde{M_2}$. We then show that $M_2$ can be implemented by a direct mechanism with refunds.

To verify that $M_2$ is a two-step mechanism, we show that $\kappa_q^v \in [0,1]$, $\tau_q^v \in [0,1]$, $\kappa_q^1 + \tau_q^1 \geq \max\left\{1, \kappa_q^0 + \tau_q^0\right\}$, and $\tau_q^0 \geq \max\left\{0, \tau_q^1\right\}$. Observe that $\kappa_q^0 + \tau_q^0 = \kappa_q^1 + \tau_q^1 = 1$, and $\tau_q^1 = 0$. Thus, we are done if we show that $\tau_q^0 \in [0,1]$. As $\widetilde{M_2}$ is a two-step mechanism, $\widetilde{\tau}_q^0 \geq \widetilde{\tau}_q^1$ and $\widetilde{\kappa}_q^1 + \widetilde{\tau}_q^1 \geq \widetilde{\kappa}_q^0 + \widetilde{\tau}_q^0$, which imply

$$\tau_q^0 = 1 + \widetilde{\tau}_q^0 - \left(\widetilde{\kappa}_q^1 + \widetilde{\tau}_q^1\right) \geq 1 + \widetilde{\tau}_q^0 - \left(\widetilde{\kappa}_q^1 + \widetilde{\tau}_q^0\right) = 1 - \widetilde{\kappa}_q^1 \geq 0; \text{ and}$$
$$\tau_q^0 = 1 + \widetilde{\tau}_q^0 - \left(\widetilde{\kappa}_q^1 + \widetilde{\tau}_q^1\right) \leq 1 + \widetilde{\tau}_q^0 - \left(\widetilde{\kappa}_q^0 + \widetilde{\tau}_q^0\right) = 1 - \widetilde{\kappa}_q^0 \leq 1.$$

$M_2$ is an IR-IC two-step mechanism because $\widetilde{M_2}$ is an IR-IC two-step mechanism, and for all $q$ and $q'$,

$$u\left(q, q'|M_2\right) - u\left(q, q'|\widetilde{M_2}\right)$$
$$= \alpha_{q'} q\left(1 - \left(\widetilde{\kappa}_{q'}^1 + \widetilde{\tau}_{q'}^1\right)\right) + \alpha_{q'}(1-q)\left(1 - \left(\widetilde{\kappa}_{q'}^1 + \widetilde{\tau}_{q'}^1\right)\right) + \widetilde{p}_q - p_q$$
$$= \alpha_{q'}\left(1 - \left(\widetilde{\kappa}_{q'}^1 + \widetilde{\tau}_{q'}^1\right)\right) - \alpha_{q'}\left(1 - \left(\widetilde{\kappa}_q^1 + \widetilde{\tau}_q^1\right)\right) = 0.$$

IR-IC two-step mechanisms $M_2$ and $\widetilde{M_2}$ are outcome-equivalent if $v(q|M_2) = v\left(q|\widetilde{M_2}\right)$ for all $q$. Observe that

$$v(q|M_2) - v\left(q|\widetilde{M_2}\right) = \alpha_q q\left(1 - \widetilde{\kappa}_q^1\right)(1 + c) + \alpha_q \times (1-q)\left(\left(\widetilde{\kappa}_q^1 + \widetilde{\tau}_q^1\right) - \left(\widetilde{\kappa}_q^0 + \widetilde{\tau}_q^0\right)\right)c$$
$$\geq 0.$$

However, $\widetilde{M_2}$ maximizes profit and therefore the above inequality must hold with equality for all $q$ in the support of $F_w$. Thus, $M_2$ is an IR-IC two-step mechanism that is outcome-equivalent to $\widetilde{M_2}$.

Lastly, for $M_2$, define a direct mechanism with refunds $M_D$ by $\alpha_0(q) = \alpha_q \kappa_q^0$, $\alpha_r(q) = \alpha_q\left(1 - \kappa_q^0\right)$, and $p(q) = p_q$. Under $M_D$, if the buyer with signal $q$ reports $q'$, then her payoff is

$$\alpha_0\left(q'\right)q + \alpha_r\left(q'\right) - p\left(q'\right) = \alpha_{q'}\kappa_{q'}^0 q + \alpha_{q'}\left(1 - \kappa_{q'}^0\right) - p_{q'}$$
$$= \alpha_{q'} q + \alpha_{q'}(1-q)\left(1 - \kappa_{q'}^0\right) - p_{q'}$$
$$= u\left(q, q'|M_2\right).$$

The seller's profit from the buyer with signal $q$ (who truthfully reports $q$) is thus

$$p(q) - \alpha_r(q)(1-q)(1+c) = p_q - \alpha_q(1-q)\left(1 - \kappa_q^0\right)(1+c)$$
$$= v(q|M_2).$$

This establishes that $M_D$ implements $M_2$.

**Proof of Lemma 4**

Consider a profit-maximizing direct mechanism with refunds $M_D = \{p(q), \{\alpha_0(q), \alpha_r(q)\}\}_{q \in [0,1]}$. If the buyer with signal $q$ reports $q'$, then her payoff is

$$U(q'; q|M_D) \equiv \left(\alpha_0^*(q') + \alpha_r(q')\right)q + \alpha_r(q')(1-q) - p(q')$$
$$= q\alpha_0(q') - \left(p(q') - \alpha_r(q')\right).$$

The incentive-compatibility condition (IC) and the individual-rationality condition (IR) are, respectively,

$$U(q; q|M_D) \geq U(q'; q|M_D) \text{ for all } q' \text{ and } q, \qquad \text{(IC)}$$
$$U(q; q|M_D) \geq 0 \text{ for all } q. \qquad \text{(IR)}$$

By applying the standard argument, we can show that the incentive-compatibility condition (IC) is equivalent to

$$\alpha_0(q) \text{ is increasing in } q \text{ and } U(q; q|M_D) = \int_0^q \alpha_0(\tilde{q})\, d\tilde{q}.$$

Thus, the seller's profit when the buyer's signal is $q$ is

$$v(q|M_D) = p(q) - \alpha_r(q)(1-q)(c+1)$$
$$= p(q) - a_r(q)\frac{1-q}{1-\gamma}$$
$$= q\alpha_0(q) + \alpha_r(q) - \int_0^q \alpha_0(\tilde{q})\, d\tilde{q} - a_r(q)\frac{1-q}{1-\gamma}$$
$$= q\alpha_0(q) + \frac{q-\gamma}{1-\gamma}\alpha_r(q) - \int_0^q \alpha_0(\tilde{q})\, d\tilde{q}.$$

Observe that $v(0|M_D) \leq 0$. Therefore, $\alpha_0(0) = \alpha_r(0) = 0$. If $q \in (0, \gamma)$, then, as $\frac{q-\gamma}{1-\gamma} < 0$, $\alpha_r(q) = 0$. If $q = \gamma$, then $v(\gamma|M_D)$ does not depend on $\alpha_r(\gamma)$. Similarly, if $q \in (\gamma, 1]$, then as $\frac{q-\gamma}{1-\gamma} > 0$, $\alpha_r(q) = 1 - \alpha_0(q)$. Summarizing these, we obtain

$$\alpha_0(q) \text{ is (weakly) increasing in } q, \quad \alpha_r(q) = \begin{cases} 0 & \text{if } q < \gamma, \\ 1 - \alpha_0(q) & \text{if } q \geq \gamma, \end{cases}$$

and,

$$v(q|M_D) \equiv \begin{cases} q\alpha_0(q) - \int_0^q \alpha_0(\tilde{q})\, d\tilde{q} & \text{if } q < \gamma, \\ q\alpha_0(q) + \frac{q-\gamma}{1-\gamma}(1 - \alpha_0(q)) - \int_0^q \alpha_0(\tilde{q})\, d\tilde{q} & \text{if } q \geq \gamma. \end{cases}$$

**Proof of Lemma 5**

We start with the case where $\gamma \leq \bar{\gamma}$. Then $F_w$ induces $q = V_\gamma^*$ with probability $1 - \gamma$ and $q = 1$ with probability $\gamma$. As $d\alpha_0(q)/dq = 0$ for all $q$ that is not in the support of $F_w$, $\alpha_0(q) = 0$ on $[0, V_\gamma^*)$, and $\alpha_0(q) = \alpha_0(V_\gamma^*)$ on $(V_\gamma^*, 1)$. Thus,

$$\mathbb{E}[v(q|a_0)]$$
$$= \Pr(q = V_\gamma^*) \times v(V_\gamma^*|M_D) + \Pr(q = 1) \times v(1|M_D)$$
$$= (1-\gamma) \times \left(V_\gamma^* \alpha_0(V_\gamma^*) + \frac{V_\gamma^* - \gamma}{1-\gamma}(1 - \alpha_0(V_\gamma^*))\right) + \gamma \times \left(1 - \alpha_0(V_\gamma^*)(1 - V_\gamma^*)\right)$$
$$= V_\gamma^*.$$

If $\gamma > \bar{\gamma}$, then the support of $F_w$ is $[V_\gamma^*, \gamma] \cup \{1\}$. Thus, $\alpha_0(q) = 0$ on $[0, V_\gamma^*)$, and $\alpha_0(q) = \alpha_0(\gamma)$ on $(\gamma, 1)$. Furthermore, $F_w$ has density $\frac{V_\gamma^*}{q^2}$ over the interval $[V_\gamma^*, \gamma)$, and two mass points, $\gamma$ (with probability $\frac{V_\gamma^*}{\gamma} - V_\gamma^*$) and 1 (with probability $V_\gamma^*$). Thus, the seller's profit from using $M_D$ is

$$\mathbb{E}[v(q|M_D)] = \Pr(q \in [V_\gamma^*, \gamma)) \times \mathbb{E}[v(q|M_D)|q \in [V_\gamma^*, \gamma)]$$
$$+ \Pr(q = \gamma) \times v(\gamma|M_D) + \Pr(q = 1) \times v(1|M_D).$$

Notice that as $F_w(q) = 1 - \frac{V_\gamma^*}{q}$ on $[V_\gamma^*, \gamma)$,

$$\Pr(q \in [V_\gamma^*, \gamma)) \times \mathbb{E}[v(q|M_D)|q \in [V_\gamma^*, \gamma)]$$
$$= \int_{V_\gamma^*}^{\gamma} \left(a\alpha_0(q) - \int_{V_\gamma^*}^{q} \alpha_0(\tilde{q}) d\tilde{q}\right) dF_w$$
$$= V_\gamma^* \int_{V_\gamma^*}^{\gamma} \frac{\alpha_0(q)}{q} dq - \int_{V_\gamma^*}^{\gamma} \left(\int_{V_\gamma^*}^{q} \alpha_0(\tilde{q}) d\tilde{q}\right) dF_w$$
$$= V_\gamma^* \int_{V_\gamma^*}^{\gamma} \frac{\alpha_0(q)}{q} dq - \left[\left(\int_{V_\gamma^*}^{q} \alpha_0(\tilde{q}) d\tilde{q}\right) F_w(q)\Big|_{V_\gamma^*}^{\gamma} - \int_{V_\gamma^*}^{\gamma} \alpha_0(q) F_w(q) dq\right]$$
$$= V_\gamma^* \int_{V_\gamma^*}^{\gamma} \frac{\alpha_0(q)}{q} dq - \left(\int_{V_\gamma^*}^{q} \alpha_0(\tilde{q}) d\tilde{q}\right)\left(1 - \frac{V_\gamma^*}{\gamma}\right) + \int_{V_\gamma^*}^{\gamma} \alpha_0(q)\left(1 - \frac{V_\gamma^*}{q}\right) dq$$
$$= -\left(\int_{V_\gamma^*}^{\gamma} \alpha_0(\tilde{q}) d\tilde{q}\right)\left(1 - \frac{V_\gamma^*}{\gamma}\right) + \int_{V_\gamma^*}^{\gamma} \alpha_0(q) dq = \frac{V_\gamma^*}{\gamma} \int_{V_\gamma^*}^{\gamma} \alpha_0(\tilde{q}) d\tilde{q}.$$

Furthermore,

$$\Pr(q = \gamma) \times v(\gamma|\alpha_0) = \left(\frac{V_\gamma^*}{\gamma} - V_\gamma^*\right) \times \left(\gamma\alpha_0(\gamma) - \int_{V_\gamma^*}^{\gamma} \alpha_0(q) dq\right), \text{ and}$$

$$\Pr(q = 1) \times v(1|\alpha_0) = V_\gamma^* \times \left(1 - \int_{V_\gamma^*}^{\gamma} \alpha_0(q) dq - \alpha_0(\gamma)(1 - \gamma)\right).$$

We thus have $\mathbb{E}[v(q|M_D)] = V_\gamma^*$.